\documentclass[%
 prl,reprint,nolongbibliography,
nofootinbib,
 amsmath,amssymb,
 aps,
floatfix,superscriptaddress
]{revtex4-2}
\usepackage{float} 
\usepackage{graphicx}
\usepackage{dcolumn}
\usepackage{bm}
\usepackage{siunitx}
\usepackage{physics}
\usepackage{xcolor}
\usepackage[caption=false]{subfig}
\usepackage[export]{adjustbox}

\usepackage{hyperref}
\hypersetup{
        colorlinks=false
}

\usepackage{nicematrix}
\NiceMatrixOptions{
code-for-first-row = \color{black} ,
code-for-last-row = \color{black} ,
code-for-first-col = \color{black} ,
code-for-last-col = \color{black}
}

\AtBeginDocument{\RenewCommandCopy\qty\SI} 
\newcommand{\aop}{\hat{a}}
\newcommand{\adag}{\hat{a}^\dagger}
\newcommand{\adagT}{\hat{a}^{\dagger 2}}

\newcommand{\sigP}{\hat{\sigma}_+}
\newcommand{\sigM}{\hat{\sigma}_-}
\newcommand{\sigZ}{\hat{\sigma}_z}
\newcommand{\sigX}{\hat{\sigma}_x}
\newcommand{\sigY}{\hat{\sigma}_y}

\begin{document}

\preprint{APS/123-QED}

\title{Multimode Qubit-Conditional Operations via Generalized Cross-Resonance}

\author{Mohammad Ayyash}
\thanks{mmayyash@uwaterloo.ca}

\affiliation{%
 Institute for Quantum Computing, University of Waterloo, 200 University Avenue West, Waterloo,
Ontario N2L 3G1, Canada}
\affiliation{Red Blue Quantum Inc., 72 Ellis Crescent North, Waterloo, Ontario N2J 3N8, Canada}

\date{\today}
    
\begin{abstract}
    
We present a general framework for generating single- and multimode qubit-conditional operations by extending cross-resonant driving to a generalized multimode scheme. This includes single-mode conditional displacements and squeezing induced by one- and two-photon cross-resonant drives in the presence of one- and two-photon qubit-oscillator interactions, respectively. In the multimode setting, we derive multimode qubit-conditional joint displacement, beamsplitter and two-mode squeezing operations. This framework enables the realization of arbitrary multimode qubit-conditional operations, which are of great importance to bosonic quantum error correction, phase estimation and quantum simulations.
\end{abstract}

\maketitle

\textit{\textbf{Introduction.---}}Bosonic quantum computation has rapidly developed as a very promising paradigm of quantum computation. In this paradigm, a logical qubit is encoded in a bosonic oscillator mode (or multiple) with the goal of suppressing one or more types of errors occurring in the oscillator \cite{Joshi_BosonicQIP2021}. Typically, these bosonic codes are then concatenated with qubit error-correcting codes such as the surface code \cite{Noh_FTBQEC2020}. With significant progress in hardware capabilities over the past decade across different implementation platforms, impressive logical state preparation has been demonstrated for various encoding schemes such as cat and Gottesman-Kitaev-Preskill (GKP) codes in superconducting circuits \cite{Reglade_CatTenSeconds2024,Eickbusch_DispUnivControl2022}, and GKP codes using trapped ions \cite{Fluhmann_TrappedIonsGKP2019}.

The workhorse and building block of the aforementioned examples is a qubit-oscillator system. Qubit-conditional operations on bosonic modes are essential for preparing logical states, reading error syndromes, and enabling universal control in the joint qubit-oscillator Hilbert space to implement logical gates \cite{Liu_HybridQOISA}. Beyond bosonic quantum error correction, qubit-conditional operations have found numerous applciations such as quantum signal processing \cite{Liu_HybridDVCVQSP_2025,Singh_nonabelianquantumsignalprocessing_2025}, quantum simulations \cite{Crane_HybridQOSims} and bosonic phase estimation \cite{Terhal_PhaseEstimation2016,Ayyash_DrivenMultiphoton2024}.

One of the most prominent qubit-oscillator operations is that of qubit-conditional rotations generated by linear qubit-oscillator interactions in the dispersive regime \cite{Blais_CircuitQED2021}. When combined with specific qubit pulses and a drive on the oscillator, the resulting operation yields a specific form of qubit-conditional displacements known as echoed-conditional displacements \cite{Terhal_PhaseEstimation2016,Eickbusch_DispUnivControl2022}. Dispersive interactions were proposed to generate conditional squeezing in the presence of a parametric two-photon drive on the oscillator \cite{VillasBoas_CondSqz_2003}. Furthermore, dispersive interactions between a qubit and two oscillators, along with a beamsplitter interaction between the two oscillators, were used to experimentally realize a qubit-conditional beamsplitter operation \cite{Mariantoni_QSwitch2009,Chapman_TunableBS2023}.

Cross-resonant driving schemes are ubiquitous in discrete-variable quantum computation with many architectures relying on cross-resonant driving between qubits to generate entangling gates such as a CNOT gate \cite{Paraoanu_CrossRes2006,Rigetti_CrossRes2010,deGroot_CrossResonanceGateExp2010,deGroot_SelectiveDarkeningTheory2012}. Cross-resonance also extends to hybrid systems composed of discrete- and continuous-variable systems. When a qubit is driven at the frequency of the oscillator, qubit-conditional displacements are generated in the oscillator \cite{Solano_CatState2003,Touzard_Readout2019,Ayyash_CatState2024}. More general cross-resonance phenomena was recently considered, where a qubit is driven at multiples of the oscillator frequency in the presence of nonlinear qubit-oscillator interactions. This results in qubit-conditional multiphoton operations such as two-photon and higher-order squeezing \cite{Ayyash_DrivenMultiphoton2024}.

In this Letter, we present a general cross-resonant driving framework applicable to generic qubit-oscillator interactions. Within this framework, we can recover single-mode qubit-conditional displacements \cite{Solano_CatState2003,Touzard_Readout2019,Ayyash_CatState2024} and squeezing \cite{Ayyash_DrivenMultiphoton2024}. In the multimode setting, we derive qubit-conditional joint displacement, beamsplitter and two-mode squeezing operations via generalized cross-resonant driving. We highlight the relevant applications for each case. We also contrast our scheme with qubit-conditional operations relying on dispersive interactions.

\textit{\textbf{Multimode qubit-conditional operations.---}}We consider a qubit coupled to many bosonic modes through an arbitrary interaction as outlined below. Additionally, the qubit is subjected to a drive. The total system Hamiltonian reads ($\hbar=1$)

\begin{subequations}
\begin{align}
    \hat{H}=\hat{H}_0 + \hat{H}_I +\hat{H}_d,
\end{align}
where
\begin{align}
    \hat{H}_0=\sum_{k}\omega_k\adag_k\aop_k + \frac{\omega_q}{2}\sigZ, 
\end{align}
\begin{align}\label{eq:IntHam}
    \hat{H}_I=\sum_{l}g_l (\sigP \hat{A}_l + \sigM\hat{A}_l^\dagger ),
\end{align}
and
\begin{align}
    \hat{H}_d=\frac{\Omega}{2}(e^{-i\omega_dt}\sigP + e^{i\omega_dt}\sigM).
\end{align}
\end{subequations}
Here, $\omega_q$ is the qubit transition frequency, $\omega_k$ is the resonance frequency of the $k$th mode, $\{g_l\}$ is the set of interaction strengths between the qubit and one or more modes specified by the interaction terms, $a_k$ ($a_k^\dagger$) is the $k$th mode's bosonic annihilation (creation) operator, $\sigZ=\dyad{e}-\dyad{g}$ represents the population difference between the qubit's ground ($\ket{g}$) and excited ($\ket{e}$) states, and $\sigM=\dyad{g}{e}$ ($\sigP=\sigM^\dagger$) is the qubit lowering (raising) operator. Furthermore, the qubit drive is characterized by strength $\Omega$ and frequency $\omega_d.$ As for the interaction terms, the operators $\hat{A}_l$ and their Hermitian conjugates, $\hat{A}_l^\dagger$, can be generic interaction operators on one or more modes (as long as the total Hamiltonian is self-adjoint). For the sake of this paper, we consider them to be powers of the creation and annihilation operators, i.e. a term of the form 
\begin{align}\label{eq:GenericIntOp}
\hat{A}_l=\aop_1^{m_{11}^{(l)}}\aop_1^{\dagger m_{12}^{(l)}}\aop_2^{m_{21}^{(l)}}\aop_2^{\dagger m_{22}^{(l)}}...\aop_N^{m_{N1}^{(l)}}\aop_N^{\dagger m_{N2}^{(l)}},
\end{align}where $N$ is the number of oscillators.

As a first step, we seek to eliminate the time dependence of the Hamiltonian by the transforming to the qubit drive's rotating frame via the unitary $\hat{U}_d=\exp(-i\omega_dt(\sum_k\adag_k\aop_k/n_k +\sigZ/2))$:
\begin{align}\label{eq:DrivRotFrameHam}
    \hat{H}^{d}&=\hat{U}_d^\dagger\hat{H}\hat{U}_d -i \hat{U}_d^\dagger \dot{\hat{U}}_d\nonumber\\ &= \sum_{k} \delta_k \adag_k\aop_k +\frac{\Delta}{2}\sigZ +\frac{\Omega}{2}\sigX \nonumber\\&\,\,\,\,\,\,\,+ \sum_l g_l( e^{i\chi_A^{(l)} t}\sigP\hat{A}_l + e^{-i\chi_A^{(l)} t}\sigM\hat{A}_l^\dagger),
\end{align}
where $\sigX=\sigP+\sigM$, $\delta_k=\omega_k-\omega_d/n_k$, $\Delta=\omega_q-\omega_d$, $\chi_A^{(l)}=\omega_d(1-f_A^{(l)})$ and $n_k$ is some positive integer dependent on the particular interaction form. Here, $f_A^{(l)}=\sum_{k}(m_{k1}^{(l)}-m_{k2}^{(l)})/n_k,$ which is obtained using the fact that the generic interaction operators are of the form in Eq.~\eqref{eq:GenericIntOp}. The choice of $n_k$'s and its relation to $f_A^{(l)}$ and $\chi_A^{(l)}$ will become apparent below.

In the drive's rotating frame, we are faced with a trivial part, a dressed qubit term and bare oscillators terms, and a nontrivial part, the interaction terms. Thus, to illuminate the dependence of the interaction term on the various parameters in the Hamiltonian, we transform to trivial part's rotating frame by means of the unitary $\hat{U}_0=\exp( -it(\sum_{k} \delta_k \adag_k\aop_k +(\Delta\sigZ +\Omega\sigX)/2))$ to obtain the interaction picture Hamiltonian
\begin{align}\label{eq:IntPicHamOrig}
    &\hat{H}^{(I)}=\hat{U}_0^\dagger\hat{H}^d\hat{U}_0 -i \hat{U}_0^\dagger \dot{\hat{U}}_0\nonumber\\=& \sum_lg_l\bigg[ \frac{\sin(\theta)}{2}\left( \dyad{\overline{+}}-\dyad{\overline{-}}\right)+\cos^2\left(\frac{\theta}{2}\right)e^{i\varepsilon t}\dyad{\overline{+}}{\overline{-}}\nonumber\\ &\,\,\,\,\,\,\,\,\,\,\,\,\,\,\,\,\,\,\,\-\sin^2\left(\frac{\theta}{2}\right)e^{-i\varepsilon t}\dyad{\overline{-}}{\overline{+}}\bigg] \hat{A}_le^{-i\Delta_A^{(l)} t} + \text{H.c.},
\end{align}
where $\Delta_A^{(l)}=\chi_A^{(l)}+\eta_A^{(l)}$, $\eta_A^{(l)}=\sum_k(m_{k1}^{(l)}-m_{k2}^{(l)})\delta_k$, and $\ket{\overline{+}}=\sin\left(\theta /2\right)\ket{{g}}+\cos\left(\theta /2\right)\ket{{e}}$ and $\ket{\overline{-}}=\cos\left(\theta /2\right)\ket{{g}}-\sin\left(\theta /2\right)\ket{{e}}$ are the dressed states with $\varepsilon=\sqrt{\Omega^2 +\Delta^2}$ and $\theta=\arctan(\Omega/\Delta)$. Note that when $\Omega\gg\Delta$, $\ket{\overline{\pm}}\simeq\ket{\pm}=(\ket{g}\pm\ket{e})/\sqrt{2}$, and when $|\Delta|\gg\Omega$, $\ket{\overline{\pm}}\simeq\ket{e/g}$. In other words, the qubit drive strength and detuning set the basis for the effective interaction.

We now briefly digress to elaborate on the relation between $n_k$'s and obtaining an \textit{interaction resonance frequency} for each interaction $\hat{A}_l.$ For each operator $\hat{A}_l,$ there exists a set of $n_k$'s such that $f_A^{(l)}=1$ and, subsequently, $\chi_A^{(l)}=0.$ 
Considering a single $\hat{A}_l$ operator and choosing an appropriate set of $n_k$'s \footnote{Explicitly, one solution is to set $n_k=|m_{k1}^{(l)}-m_{k2}^{(l)}|.$} that ensure $f_A^{(l)}=1$, we can define the \textit{interaction resonance frequency} of $\hat{A}_l$ as
\begin{align}
    \omega_A^{(l)}=\left|\sum_{k}n_k\omega_k(m_{k1}^{(l)}-m_{k2}^{(l)})\right|.
\end{align}
Generally, for two (or more) operators, $\hat{A}_l$ and $\hat{A}_m$ with $l\neq m$, the set of $n_k$'s that impose $f_A^{(l)}=1$ and $f_A^{(m)}=1$ are different. In this paper, we restrict our attention to the special case where a single set of $n_k$'s ensures that $f_A^{(l)}=1$ for all $l.$ This condition's physical meaning is that the polynomial degree in all creation and annihilation operators for all $\hat{A}_l$ must be the same, i.e. the integer $\sum_k(m_{k1}^{(l)}+m_{k2}^{(l)})$ is the same for all $l.$ This means that all $\hat{A}_l$ terms have the same interaction resonance frequency. Hence, since we will only consider cases where all $\hat{A}_l$'s have the same interaction resonance frequency, we simply drop the superscript $(l)$ and denote it $\omega_A$. We do the same thing for $f_A^{(l)},\,\chi_A^{(l)}$ and $\Delta_A^{(l)}.$ Then, this simplifies things as $\Delta_A$ is the same for all $l,$ and it is the detuning between the qubit drive and interaction resonance. When $\Delta_A=0$, we have $\omega_d=\omega_A$, which makes the qubit drive \textit{cross-resonant} with the interaction. This serves as a generalized cross-resonance extending the case from driving at some positive integer multiple of the oscillator frequency.

The terms within the square bracket of Eq.~\eqref{eq:IntPicHamOrig} consist of two sets of interactions occurring on separate timescales; a set of the interactions oscillates with $e^{i\pm\varepsilon t}$. Thus, when the driving amplitude, $\Omega$, and/or the qubit detuning, $\Delta$, are significantly larger than $\Delta_A$ and $\,g$, we can eliminate the fast-oscillating terms by imposing the rotating-wave approximation (RWA) via the condition
\begin{align}\label{eq:GenCrossResRWA}
    |\Delta_A|,\,g\ll\varepsilon.
\end{align}
This condition imposes that $\varepsilon$ is the largest energy scale and yields the effective Hamiltonian
\begin{align}
    \hat{H}_{\text{Eff}}=(\dyad{\overline{+}}-\dyad{\overline{-}})\sum_l(\overline{g}_l(t)\hat{A}_l + \overline{g}_l^*(t)\hat{A}_l^\dagger),
\end{align}
where $\overline{g}_l(t)=e^{i\Delta_At}g_l\sin(\theta)/2$ and $\overline{g}^*(t)$ is the complex conjugate of $\overline{g}(t).$ This last Hamiltonian serves as the basis for this study. It conditions the oscillator(s) interactions, $\sum_l\overline{g}_l(t)\hat{A}_l + \overline{g}_l^*(t)\hat{A}_L^\dagger,$ on the qubit state. Its effective time-evolution operator reads
\begin{align}
    \hat{U}(t,0)=&\dyad{\overline{+}}\mathcal{T}\left\{e^{-i\int_0^t d\tau \hat{H}_{\text{Eff}}(\tau)}\right \}\nonumber\\&+\dyad{\overline{-}}\mathcal{T}\left\{e^{+i\int_0^t d\tau \hat{H}_{\text{Eff}}(\tau)}\right \}
\end{align}
with $\mathcal{T}\{.\}$ being the time-ordering operator. Generally, we restrict our attention to cases where the effective Hamiltonian is time independent, which occurs when the qubit drive is near generalized cross-resonance, i.e. $\omega_d\simeq\omega_A$ ($\Delta_A\simeq0$). In this case, the time-ordering is unnecessary, and the time-evolution operator becomes
\begin{align}\label{eq:EffTimeEvOp}
    \hat{U}(t,0)=\dyad{\overline{+}}e^{-i\int_0^t d\tau \hat{H}_{\text{Eff}}(\tau)}+\dyad{\overline{-}}e^{+i\int_0^t d\tau \hat{H}_{\text{Eff}}(\tau)}.
\end{align}
A simple and intuitive interpretation of the time-evolution operator's form is the following: the direction of time (forward or backward) is conditioned on the qubit state, $\ket{\overline{+}}$ or $\ket{\overline{-}},$ due to the relative phase imposed by the particular qubit state.

\textit{\textbf{Single-mode examples.---}}Let us specify two established single-mode examples that will pave the way towards generalizing to multimode cases. The first example is one where the qubit-oscillator interaction Hamiltonian is the Jaynes-Cummings (JC) interaction, $\hat{H}_I=g(\sigP\aop+\sigM\adag)$ \footnote{In fact, this will also work with the anti-JC interaction, $\sigP\adag+\sigM\aop.$}. In this case, there is only one oscillator and one $\hat{A}_l$ operator; we have $m_{11}^{(1)}=1$ and $m_{12}^{(1)}=0$, which require $n_1=1$ such that $f_A=1$ and $\chi_A=0$. Here, the interaction resonance frequency is simply the oscillator's frequency, $\omega_A=\omega_1$. Imposing the cross-resonance condition, $\omega_d=\omega_A$, yields qubit-conditional displacements, where Eq.~\eqref{eq:EffTimeEvOp} becomes \cite{Ayyash_CatState2024,Solano_CatState2003}
\begin{align}
    \hat{U}(t,0)=\dyad{\overline{+}}\hat{D}(\alpha(t))+\dyad{\overline{-}}\hat{D}(-\alpha(t))
\end{align}
with $\hat{D}(\alpha)=e^{\alpha\adag-\alpha^*a}$ being the displacement operator and $\alpha(t)=-igt/2.$ Cross-resonant conditional displacements were experimentally used for qubit readout \cite{Touzard_Readout2019}. Additionally, qubit-conditional displacements are necessary for universal control over the joint qubit-oscillator Hilbert space, which enables the preparation of logical bosonic codewords such as GKP states \cite{Eickbusch_DispUnivControl2022,Weigand_BreedingGridStates2018}. 

The second single-mode example relies on a two-photon JC interaction, $\hat{H}_I=g(\sigP\aop^2+\sigM\adagT)$. In this case, we have $m_{11}^{(1)}=2$ and $m_{12}^{(1)}=0$, which requires $n_1=2$ to ensure that $f_A=1$ and $\chi_A=0.$ The interaction resonance frequency, in this case, is $\omega_A=2\omega_1.$ Similarly to the one-photon interaction above, driving the qubit at $\omega_A$ (two-photon cross-resonance) results in qubit-conditional squeezing, where Eq.~\eqref{eq:EffTimeEvOp} becomes \cite{Ayyash_DrivenMultiphoton2024}
\begin{align}
    \hat{U}(t,0)=\dyad{\overline{+}}\hat{S}(\zeta(t))+\dyad{\overline{-}}\hat{S}(-\zeta(t))
\end{align}
with $\hat{S}(\zeta)=e^{(\zeta^*\adagT-\zeta \aop^2)/2}$ being the squeezing operator and $\zeta(t)=igt.$ Qubit-conditional squeezing can be used in bosonic phase estimation \cite{Ayyash_DrivenMultiphoton2024}. Additionally, it can facilitate faster universal control since it can generate higher-order polynomials in the Lie algebra of polynomial operators on the joint qubit-oscillator Hilbert space, i.e. creating generators such as $\hat{\sigma}_j\hat{x}^r\hat{p}^s$ using a lower number of steps (or reduced circuit depth) with $\hat{\sigma}_j\in\{\sigX,\sigY,\sigZ\}$, $\hat{x}=(\adag+\aop)/\sqrt{2}$ and $\hat{p}=i(\adag-\aop)/\sqrt{2}$ being generalized position and momentum quadrature operators for some non-negative integers $r$ and $s$ \cite{Ayyash_DrivenMultiphoton2024,Liu_HybridQOISA}. This unitary operator can also produce superpositions of opposite-phase squeezed states (after an appropriately selected qubit measurement), which happen to have various unique photon-number interference properties that make them potentially useful for encoding logical qubits in the oscillator \cite{Ayyash_DrivenMultiphoton2024,DelGrosso_CtrlSqzGate2024}. This operation has already been experimentally utilized in a trapped ion setup, although it relies on a completely different mechanism to generate the conditional squeezing \cite{Saner_NonclassicalQHOStates2024}.

\textit{\textbf{Multimode examples.---}}With the single-mode examples of one- and two-photon cross-resonance specified within this general framework we outlined, we now move on to considering multimode cases with their generalized notion of cross-resonance. We focus on three key examples: qubit-conditional joint displacement, beamsplitter and two-mode squeezing operations.

We now consider a qubit simultaneously interacting with many modes via a linear JC interaction such that the interaction Hamiltonian is $\sum_lg_l(\sigP\aop_l+\sigM\adag_l)$. In this case, we have $m_{k1}^{(l)}=1$ for all $k$ and $l$ with all other $m_{kj}^{(l)}=0$. Then, we require $n_k=1$ for all $k$, and, as such, the interaction resonance, in this case, is $\omega_A=\sum_k \omega_k$. When the qubit is driven at $\omega_A$, the resulting time-evolution operator is that of a qubit-conditional joint displacement on all modes
\begin{align}
    \hat{U}(t,0)=\dyad{\overline{+}}\prod_{l}\hat{D}_l(\alpha_l(t))+\dyad{\overline{-}}\prod_l\hat{D}_l(-\alpha_l(t)),
\end{align}
where the displacements $\alpha_l(t)=-ig_lt/2$. These multimode conditional displacements are needed for the preparation of logical states and syndrome measurements in multimode GKP codes \cite{Gottesman_Oscillator2001,Royer_MultimodeOsc2022}. In practice, this operation can be easily realized for two quasi-resonant modes $(\omega_1\simeq\omega_2)$ such as two orthogonal modes in a box cavity. This operation would still work even if the modes are not quasi-resonant \footnote{See Ref.~\cite{Ayyash_CatState2024} for details on the cases deviating away from exact cross-resonance.} such as in Ref.~\cite{Diringer_CNOD2024} or in the case of multi-post multimode cylindrical cavities \cite{Lemonde_BlueprintNordQuantique2024}. In such scenarios, one can place the qubit drive frequency halfway between both modes such that $|\delta_1|\simeq|\delta_2|$. Unlike multimode echoed-conditional displacements \cite{Eickbusch_DispUnivControl2022,You_MultimodeControl2024}, which need a drive on each oscillator along with properly timed pulses on the qubit, our method only needs a single qubit drive. With echoed-conditional displacements, the amplitudes of the drives in the dispersive regime lead to spurious rotations and qubit dephasing \cite{Eickbusch_DispUnivControl2022}. Cross-resonant joint displacements do not suffer from this. However, in our proposal, the amplitudes of the displacements are limited by the qubit-oscillator interaction strengths, $g_k$. Additionally, cross-resonant joint displacements will simultaneously displace all oscillators, which can be easily corrected with unconditional displacements at the end of the operation \cite{Weigand_BreedingGridStates2018,Eickbusch_DispUnivControl2022}.

Next, we turn our attention to the case of two modes interacting with the qubit via $\hat{H}_I=g(\sigP\aop_1\adag_2+\sigM\adag_1\aop_2)$. In this example, $m_{11}^{(1)}=m_{22}^{(1)}=m_{12}^{(2)}=m_{21}^{(2)}=1$ with all other $m_{kj}^{(l)}=0.$ Here, we require $n_1=n_2=1$ such that $f_A=1.$ Then, the interaction resonance is $\omega_A=|\omega_1-\omega_2|.$ When the qubit is driven at $\omega_A$, the time-evolution operator for this case yields a qubit-conditional beamsplitter operation
\begin{align}
    \hat{U}(t,0)=\dyad{\overline{+}}\hat{B}_{12}(\theta(t),\phi(t))+\dyad{\overline{-}}\hat{B}_{12}(-\theta(t),\phi(t)),
\end{align}
where $\hat{B}_{12}(\theta,\phi)=e^{\theta(e^{i\phi}\aop_1\adag_2-e^{-i\phi}\adag_1\aop_2)}$ is the beamsplitter operator with $\theta(t)=|-igt/2|$ and $\phi(t)=\arg(-igt/2)$. Interestingly, the beamsplitter interaction resonance frequency is the detuning between the oscillators. This operation can be used as a qubit-controlled `router' of photons between the two modes \cite{Mariantoni_QSwitch2009}. Additionally, this operation is useful for gates and readout in bosonic dual-rail qubits \cite{Gan_CondBSGate_2020,deGraaf_CondBSMidCheck_2025}.

Another interesting case is that where the qubit-two-mode interaction is $\hat{H}_I=g(\sigP\aop_1\aop_2+\sigM\adag_1\adag_2)$. Following the case of the qubit-conditional beamsplitter, we find $m_{11}^{(1)}=m_{21}^{(1)}=m_{12}^{(2)}=m_{22}^{(2)}=1$ with all other $m_{kj}^{(l)}=0$ and require that $n_1=n_2=1$. Consequently, the interaction resonance frequency is $\omega_A=\omega_1+\omega_2.$ Thus, driving the qubit at $\omega_A$, we find a qubit-conditional two-mode squeezing operation
\begin{align}
    \hat{U}(t,0)=\dyad{\overline{+}}\hat{S}_{12}(\zeta(t))+\dyad{\overline{-}}\hat{S}_{12}(-\zeta(t)),
\end{align}
where $\hat{S}_{12}(\zeta)=e^{\zeta\adag_1\adag_2-\zeta^*\aop_2\aop_2}$ is the two-mode squeezing operator with $\zeta(t)=-igt/2$. Qubit-conditional two-mode squeezing can be used for quantum simulations of relativistic quantum field effects involving photon pair creation, such as the Unruh effect \cite{Hu_UnruhEffectSim2019}. This can be generalized to many oscillators via the interaction Hamiltonian $\hat{H}_I=\sum_{l,m}g_{l,m}(\sigP\aop_l\aop_m+\sigM\adag_l\adag_m).$

\textit{\textbf{Discussion.---}} While we focused on qubit-conditional Gaussian operations in our examples, it is worth noting that the framework also works for more general qubit-conditional non-Gaussian operations. For example, if we consider a qubit interacting with two modes by means of some generalized multiphoton spontaneous parametric down-conversion (or up-conversion) interaction Hamiltonian $\hat{H}_I=g(\sigP\aop_1^{m_1}\aop_2^{\dagger m_2}+\sigM\aop_1^{\dagger m_1}\aop_2^{m_2})$, then by driving the qubit at the appropriate interaction resonance frequency, we can obtain a qubit-conditional operation of the form 
\begin{align}\label{eq:GenericQCOp}
    \hat{U}(t,0)=\dyad{\overline{+}}\hat{V}_{}(\zeta(t))+\dyad{\overline{-}}\hat{V}(-\zeta(t))
\end{align}
where $\hat{V}(\zeta(t))=e^{\zeta(t)\aop_1^{m_1}\aop_2^{\dagger m_2}-\zeta^*(t)\aop_1^{\dagger m_1}\aop_2^{m_2}}.$ Thus, the framework extends well beyond qubit-conditional Gaussian operations.

In addition to some of the aforementioned applications, general multimode qubit-conditional operations, such as that in Eq.~\eqref{eq:GenericQCOp}, are useful for bosonic phase estimation \cite{Terhal_PhaseEstimation2016,Ayyash_DrivenMultiphoton2024}, multimode qubit-oscillator quantum signal processing \cite{Liu_HybridDVCVQSP_2025,Singh_nonabelianquantumsignalprocessing_2025} and universal control over joint qubit-multimode Hilbert spaces \cite{You_MultimodeControl2024,Liu_HybridQOISA}.

The necessary multimode qubit-oscillator interactions, $\sigP\hat{A}_l+\sigM\hat{A}_l^\dagger$, for all the different cases we considered can be readily realized using superconducting circuits by coupling, for example, a transmon qubit to a multimode resonator (planar or three-dimensional) or many single-mode resonators by means of a nonlinear coupler such as an asymmetric superconducting quantum interference device (SQUID) \cite{Ayyash_DrivenMultiphoton2024,Stolyarov_TwoPhSQUID_2025,Stolyarov_TwoPhSQUID2_2025} or a superconducting nonlinear asymmetric inductive element (SNAIL)  \cite{Zorin_Snail2016,Frattini_Snail2017}. Experiments relying on native three-body interactions such as the ones we use in this work have already been realized \cite{Chang_ThreePhSPDC2020,Busnaina_NativeThreeBodyInt2025}. For the case of qubit-conditional Gaussian operations, all of these can be readily implemented in a state-of-the-art device using a nonlinear coupler as they are at most third-order interactions \footnote{For example, the requisite interactions for qubit-conditional single- and two-mode squeezing are $\sigP\aop^2+\sigM\adagT$ and $\sigP\aop_1\aop_2+\sigM\adag_1\adag_2$, respectively, which are both third-order interactions.}. However, for more general qubit-conditional non-Gaussian operations as in Eq.~\eqref{eq:GenericQCOp}, this would depend on the order of the interactions needed to generate such a unitary and the coupling strength compared to spurious effects arising \cite{Chang_ThreePhSPDC2020}.

It is worth highlighting the differences between our proposed generalized cross-resonance qubit-conditional operations and those relying on dispersive interactions. First, let us highlight that a qubit-multimode system in the dispersive regime with an effective interaction Hamiltonian
\begin{align}
    \hat{H}_I=\sum_k\xi_k\sigZ(\adag_k\aop_k) +\sum_l g_l(t)(\hat{A}_l^\dagger +\hat{A}_l),
\end{align}
can effectively realize approximate qubit-conditional $\hat{W}(\vec{\gamma}(t))=\exp(\sum_l(\gamma_l(t)\hat{A}_l-\gamma_l^*(t)\hat{A}^\dagger_l))$ operations \cite{Eickbusch_DispUnivControl2022,You_MultimodeControl2024,Chapman_TunableBS2023}. Here, $\xi_k$ is the dispersive shift of each mode and $\vec{\gamma}=(\gamma_1,...,\gamma_M)$ is a complex vector with $M$ being the number of interaction terms. The reliance on the dispersive interactions for generating qubit-conditional operations is restricted by the dispersive regime's perturbative bounds \cite{Strauch_ArbControl_2010,Chapman_TunableBS2023}. Additionally, the desired interactions which we seek to make conditioned on the qubit state, $\sum_lg_l(\hat{A}_l^\dagger+\hat{A}_l)$, must be time dependent in this scheme \cite{Eickbusch_DispUnivControl2022,Chapman_TunableBS2023}. Spurious cross-Kerr terms arise due to dispersive interactions and affect the fidelity of the effective qubit-conditional operation \cite{You_MultimodeControl2024,Chapman_TunableBS2023}. These undesired terms do not arise in our scheme. However, our scheme is limited by the qubit-oscillator interaction strengths as well as the RWA condition of Eq.~\eqref{eq:GenCrossResRWA}, which can result in shifts in the qubit and oscillator frequencies that can also decrease the fidelity of the qubit-conditional operations \cite{DFV_EffectiveHamTheory2007,Ayyash_CatState2024}. Another disadvantage of our scheme is the need for native multi-body interactions between the qubit and oscillators as in Eq.~\eqref{eq:IntHam}, which are usually realized with weak coupling strengths \cite{Chang_ThreePhSPDC2020,Busnaina_NativeThreeBodyInt2025}. This consequently limits the amplitude of the operation parameter such as the displacement amplitude, squeezing amplitude or beamsplitter angle. One advantage our scheme possesses is the flexibility in the parameter regimes. The cross-resonant framework can operate outside the dispersive regime, and there is a tradeoff between the qubit drive strength, $\Omega$, and qubit detuning, $\Delta$, that can be utilized \cite{Ayyash_DrivenMultiphoton2024}.

The framework developed here can also be extended to the case of qudit-oscillator interactions. An instance of such an extension was studied in Ref.~\cite{Ayyash_CatState2024} where qubit-conditional displacements were extended to the case of a qutrit. Future work could generalize our framework for qudits with particular selection rules or allowed transitions that could arise in a realistic implementation. 

\textit{\textbf{Conclusions.---}} We developed a framework that generalizes cross-resonant driving schemes to multimode qubit-oscillator interactions yielding qubit-conditional operations. We highlighted examples of such conditional operations including joint conditional displacement, beamsplitter and two-mode squeezing operations. Additionally, we contrasted our approach to existing approaches relying on dispersive interactions to generate the desired qubit-conditional operations. This framework is simple yet power as it enables the realization of arbitrary multimode qubit-conditional operations, which are instrumental to bosonic quantum error correction \cite{Liu_HybridQOISA}, phase estimation \cite{Terhal_PhaseEstimation2016,Ayyash_DrivenMultiphoton2024} and universal multimode control \cite{You_MultimodeControl2024,Singh_nonabelianquantumsignalprocessing_2025}. We believe that our proposal offers a new avenue for multimode qubit-conditional operations that will be of great value to near-term experimental efforts in qubit-oscillator systems.

\textit{\textbf{Acknowledgements.---}}The author thanks Sahel Ashhab, Xicheng (Christopher) Xu, Jamal H Busnaina, Matteo Mariantoni and Noah Gorgichuk for useful discussions.  This work was supported by the Institute for Quantum Computing (IQC) through funding provided by Transformative Quantum Technologies (TQT).

\appendix
\bibliography{paper}

\end{document}